\documentclass[fleqn,10pt]{wlscirep}
\pdfoutput=1
\usepackage{hyperref}
\usepackage{epsfig}
\usepackage{amsmath}
\usepackage{graphicx}
\usepackage{wrapfig}
\usepackage{dcolumn}
\usepackage{bm}
\usepackage{amssymb}
\usepackage{titlesec}
\usepackage{mathtools}
\usepackage{relsize}
\usepackage{cleveref}
\usepackage{cite}

\title{Large order fluctuations, switching, and control in complex networks}

\author[1*]{Jason Hindes}
\author[1]{Ira B. Schwartz}
\affil[1]{U.S. Naval Research Laboratory, Code 6792, Plasma Physics Division, Nonlinear Systems Dynamics Section, Washington, DC 20375}

\affil[*]{jason.hindes.ctr@nrl.navy.mil}

\begin{abstract}
We propose an analytical technique to study large fluctuations and switching from internal noise in complex networks. Using order-disorder kinetics as a generic example, we construct and analyze the most probable, or optimal path of fluctuations from one ordered state to another in real and synthetic networks. The method allows us to compute the distribution of large fluctuations and the time scale associated with switching between ordered states for networks consistent with {\it mean-field} assumptions.  In general, we quantify how network heterogeneity influences the scaling patterns and probabilities of fluctuations. For instance, we find that the probability of a large fluctuation near an order-disorder transition decreases exponentially with the {\it participation ratio} of a network's principle eigenvector -- measuring how many nodes effectively contribute to an ordered state. Finally, the proposed theory is used to answer how and where a network should be targeted in order to optimize the time needed to observe a switch.
\end{abstract}
\begin{document}

\flushbottom
\maketitle
%
%
\thispagestyle{empty}

\section{Introduction}
Network science is highly interdisciplinary, and when combined with other fields such as statistical physics and nonlinear dynamics, provides a useful framework to address fundamental questions regarding complex systems\cite{Dorogovtsev:RMP2008,Vespignani1:Book,Newman:Book}. Network approaches have provided quantifiable results in diverse applications in nearly every field of science and engineering, from biological networks\cite{ProulxRev2005} to climate networks\cite{Berezin2012}, information networks\cite{Timar2017}, infrastructure networks\cite{Bagler2008}, and social networks\cite{Weng2013}. Consequently, much progress has been made in understanding the role of topology in many collective processes in complex systems, including in adaptive and co-evolving networks, where the topological dynamics is itself a function of a network's state \cite{Gross:Book}. Popular examples where interaction structure is understood to strongly influence behavior are the spread of infectious diseases \cite{PastorRMP}, the dynamics of neural systems \cite{Wang2013}, the synchronization of coupled oscillators\cite{Lou2014}, the patterns of voters\cite{Gracia2014}, and the collective motion of networked swarms \cite{HindesSwarm2016}. 

However, many theoretical results in network dynamics rely on deterministic, and ``mean field" limits of some simple model\cite{Vespignani1:Book}. Though useful, such approaches typically ignore noise and dynamical fluctuations that are inherent in virtually all of the aforementioned examples. Although inherent noise may be considered small in large networks, the existence and observation of large fluctuations can result in drastic change in a network's dynamics\cite{HindesExtin2016}. Therefore, some recent efforts have been made to study the relationship between network dynamics and noise\cite{Carro2016,Ching2017,Havlin2017}. It has been demonstrated that the interplay between complex topology and noise can alter well-known scaling laws and patterns for fluctuations \cite{Assaf2012}, as well as provide new control mechanisms that take advantage of noise-induced phenomena including switching and extinction  \cite{HindesExtin2016,Motter2015}. 

An important class of statistical physics models for capturing many processes, and where noise is relevant, are spin systems, in which nodes in a network take on discrete states\cite{Dorogovtsev:RMP2008,Vespignani1:Book}. Typically, the probability (or probability of changing in time) of any configuration of states depends on the configuration's energy \cite{Castellano2009}. Such an approach has been useful for understanding opinion formation and dynamics \cite{Castellano2006,Dorogovtsev2002}, rumor spreading\cite{Ostilli2010}, as well for machine learning on networks (e. g., Boltzmann machines)\cite{Tanaka1998} and network inference\cite{Zeng2010}. In spin dynamics, often two limits are considered. The first entails zero temperature, where the energy of a network is minimized and generally does not fluctuate between configurations\cite{Sanders2009,Castellano2006}. The second involves the average state of a network at finite temperatures and noise, where an order-disorder transition is observed and analyzed in the limit where the network size tends to infinity \cite{Dorogovtsev2002}. However, all real networks are finite systems. Thus, fluctuations between distinct metastable configurations arise -- effectively changing the collective order due to noise. Example systems are social networks, where fluctuations are known to be important\cite{Xiong2014,Ace2013} and hence switching from one majority opinion to another is possible. Yet, many open questions remain about how switching occurs in complex networks as a result of random fluctuations.

On the other hand, the role of noise is reasonably well understood in simple well-mixed and spatially homogenous contexts \cite{Dykman1994,Meerson2010,Kamenev2008,AssafRev}. It has been demonstrated in many works that noise and collective dynamics can couple in such a way as to induce a large fluctuation -- effectively driving a system to switch from one collective behavior to another.  If the fluctuation is a { \it rare event}, then the process is captured by a most probable, or optimal path (OP) -- where all others are exponentially less likely to occur \cite{Assaf2010,Lindley2014}. In such cases, the OP is describable in an analytical-mechanics formalism
familiar from classical physics \cite{Dykman1994}, which has been used to elegantly describe a variety of rare phenomena including: fixation in evolutionary games\cite{MobiliaEPL2010}, extinction of disease in homogenous populations\cite{SchwartzJST2009}, switching in self-regulating genes \cite{Assaf2011}, large velocity fluctuations in propagating fronts\cite{MeersonPRER2011}, viral clearance\cite{Chaudhury2012}, irreversible fluctuations in electronic circuits\cite{Luchinsky1997}, and switching in quantum mechanical oscillators\cite{Lin2015}. Our strategy is to find the OP in general network configurations given a general spin (opinion) dynamics, and use it to understand the dynamical switching pathway between ordered (majority) states, the average time needed to see a switch, and the distribution of large fluctuations in both real and synthetic networks.  

This report moves well beyond both the deterministic and small fluctuation limits of order-disorder dynamics in complex networks by addressing how large fluctuations occur. 
Our approach enables us to construct and analyze the approximate OP through a given network -- reducing a very high dimensional stochastic and rare process to a single trajectory. Consequently, we are able to compute several topologically dependent quantities that have so far eluded analysis in network science: the distribution, shape, and time scale of large fluctuations. Beyond computation, we show that just above an order-disorder transition, the probability of a large fluctuation decreases exponentially with the participation ratio of a network \cite{Pastor2016}, and hence is exponentially sensitive to topological heterogeneity-- e.g., the ratio of second to fourth moments of a network's degree distribution. Moreover, we find two quantitatively distinct scaling patterns for the distribution of large fluctuations, in which low eigenvector-centrality fluctuations predominate at high order (large majorities), and high centrality at low order (small majorities).  
Finally, we demonstrate with several examples on a Facebook network\cite{Mcauley2012} how the formalism is useful for designing controls that optimally leverage noise in order to minimize the time scale for switching -- answering where and at what rate a network should be targeted in order to induce a switch.   


\section{Methods}
\subsection{Model definition and mean-field dynamics}
We consider a system of $N$ nodes interacting through a network. The network is represented by a real-valued matrix, A, where $A_{ij}$ gives the influence strength of node
$i$ on node $j$. At any instant, each node $i$ is in one of two possible opinion (spin) states, characterized by $s_{i}\!\in\!\{-1,1\}$ \cite{Castellano2009,Carro2016}. Nodes can change state by interacting with their nearest neighbors in the network, such that $s_{i}$ evolves {\it stochastically} in time with a probability that depends on $s_{i}$ and $s_{j}$ of neighbors, for $A_{ij}\neq0$. We study a simple interaction rule motivated from statistical physics, where each node has a tendency to align its opinion with its neighbors such that the energy, $E_{i}\!\equiv\!-s_{i}\sum_{j}A_{ij}s_{j}$, is minimized \cite{Vespignani1:Book,Castellano2009,Ben-Naim:Book}. We choose the kinetics to be a continuous-time Glauber dynamics-- a Markov process with a transition rate for each node:
\begin{align}
\text{Rate}(s_{i}\!\rightarrow\!-s_{i})=\frac{\alpha}{1+e^{-2\beta E_{i}}}\!+\!f_{i},
\label{eq:Flip}
\end{align}
where $\beta$ is an inverse temperature that measures the ratio of energy to thermal noise, $f_{i}$ is a local spontaneous flipping rate \cite{Carro2016}, and $\alpha$ is a rate constant that determines the units of time; we take $\alpha\!=\!1$ without loss of generality. Qualitatively, minimizing energy pushes the network toward complete order (unanimous majority or consensus), while thermal noise and spontaneous flipping inject fluctuations that tend to break up order. Because of this interplay, a majority order emerges as long as $\beta$ is above some critical value (see SI.\ref{sec:2} for derivation)\cite{Lynn2016,Dorogovtsev2002}. Below, our results are to be compared with the stochastic process defined by Eq.(\ref{eq:Flip}). 

In general, the network dynamics is governed by a master equation for the $N$-node probability distribution, $\rho(\bold{s},t)$, which is high-dimensional and difficult to analyze in its entirety. Therefore, we seek a simplified description of the dynamics by first considering the opinion {\it density} at each node in a large ensemble. The ensemble consists of $C$ identical networks with the same $A$, but {\it independent} realizations of the dynamics given by Eq.(\ref{eq:Flip}). The opinion density in the ensemble is $m_{i}\!\equiv\!\sum_{c=1}^{C}s_{i,c}/C$, where $s_{i,c}$ is the state of node $i$ in realization $c$ of the stochastic process. Ultimately, we are interested in the limit $C\rightarrow \infty$, or continuous density. Our goal is to find an approximate master equation for the network ensemble distribution, $P(\bold{m},t)$, that is a function of the densities alone, and extract a particular solution relevant to large fluctuations. As we will see, such a solution will correspond to the OP. To find it, we must consider the transition rates for $m_{i}$ and make a {\it mean-field} approximation: 
\begin{align}
R_{i}^{\pm}(\bold{m})\!\equiv\!\text{Rate}(m_{i}\!\rightarrow\!m_{i}\pm2/C)=\sum_{c}\frac{1}{2}(1\mp s_{i,c})\!\!\Bigg[\frac{1}{1+e^{\mp2\beta\sum_{j}A_{ij}s_{j,c}}}\!+f_{i}\Bigg]\approx\frac{C}{2}(1\mp m_{i})\!\!\Bigg[\frac{1}{1+e^{\mp2\beta\sum_{j}A_{ij}m_{j}}}\!+f_{i}\Bigg]. 
\label{eq:MFrate}
\end{align}
The mean-field approximation replaces $s_{i,c}$ by its ensemble average $m_{i}$ -- effectively neglecting correlations between neighbors in the network. The result is a master equation that describes a simplified stochastic process in terms of the opinion density at each node,
\begin{align}
\label{eq:MasterEquation}
&\frac{\partial P}{\partial t}(\bold{m},t)=\sum_{i}R^{+}_{i}(\bold{m}\!-\!\frac{2}{C}\bold{1}_i)P(\bold{m}\!-\!\frac{2}{C}\bold{1}_i,t)-R^{+}_{i}(\bold{m})P(\bold{m},t)+R^{-}_{i}(\bold{m}\!+\!\frac{2}{C}\bold{1}_i)P(\bold{m}\!+\!\frac{2}{C}\bold{1}_i,t)-R^{-}_{i}(\bold{m})P(\bold{m},t),
\end{align}
where $\bold{1}_{i}=\left<0{}\;_{1},0{}\;_{2},...,0{}\;_{i-1},1{}\;_{i}, 0{}\;_{i+1},...\right>$. In Sec.\ref{sec:results} we analyze Eq.(\ref{eq:MasterEquation}) and compare to simulations of Eq.(\ref{eq:Flip}) on several networks.   

\section{\label{sec:results}Results}
\subsection{Large fluctuations}
\begin{figure}[ht]
\centering
\includegraphics[scale=0.35]{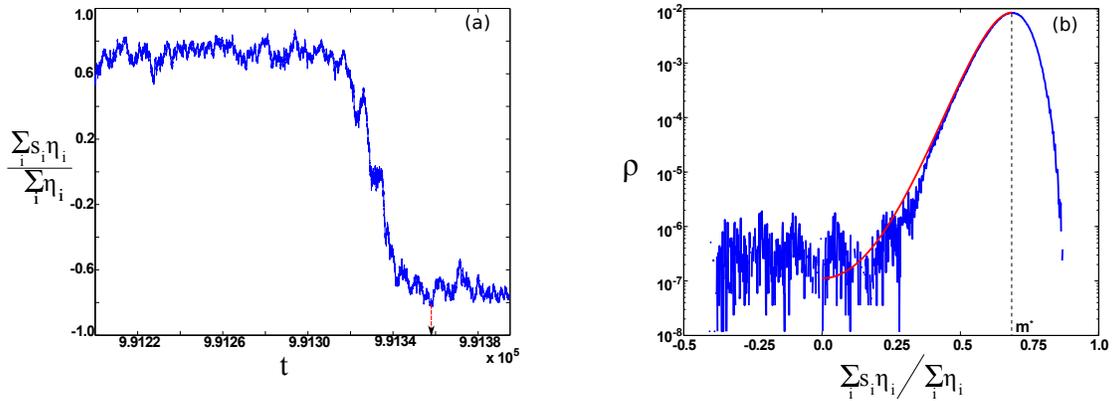}
\caption{(a) Switching in a Facebook network\cite{Mcauley2012} between meta-stable ordered states. Average opinion, weighted by eigenvector centrality ($\eta_{i}$ for node $i$), is shown versus time. Parameters are $\beta\lambda\!=\!1.37$ and $f_{i}\!=\!0.02\;\forall i$, where $\lambda$ is the largest eigenvalue of $A$. The arrow indicates the switching time. (b) Histogram of the stochastic trajectory corresponding to (a). The predicted distribution is shown in red from solving Eqs.(\ref{eq:Action}-\ref{eq:EOM2}).}
\label{fig:Switch}
\end{figure}

When $\beta$ is above threshold, $P(\bold{m},t)$ is peaked around one of two ordered equilibrium states, $\bold{m}(t)\approx\pm\bold{m}^*$. Dynamically, a finite network fluctuates around equilibrium, $\bold{m}^*$, after an initial transient, for a long time until a large fluctuation occurs, which carries the network to the opposite ordered state, $-\bold{m}^*$, as shown in Fig.\ref{fig:Switch}(a). Such order switches are rare events in large networks ($N\!\gg\!1$), and we expect them to be encoded in the tails of $P(\bold{m},t)$. In particular, if $\bold{m}$ corresponds to a large deviation from $\bold{m}^*$, we expect an exponential reduction in probability, as demonstrated in Fig.\ref{fig:Switch}(b). Therefore, it is convenient to constrain our search for solutions of the master equation to the exponentially-distributed tail that is relevant for rare events, since Eq.(\ref{eq:MasterEquation}) contains too much information to be useful in practice. 


These observations suggest extracting a solution of Eq.(\ref{eq:MasterEquation}) with exponential, or Wentzel--Kramers--Brillouin $(WKB)$ form, $P(\bold{m},t)\!=\!ae^{-CS(\bold{m},t)}$. The WKB solution for the ensemble distribution can be viewed as a product of independent and identical distributions for each realization in the ensemble. Thus, we can approximate the probability distribution for states in a single realization, $\rho(\bold{s},t)$, by
\begin{align}
\label{eq:Distribution}
\rho(\bold{s},t)\cong\rho(\bold{m},t)=be^{-S(\bold{m},t)}. 
\end{align}
Predictions from Eq.(\ref{eq:Distribution}) are in good agreement with simulations on a real Facebook network\cite{Mcauley2012} -- shown in red in Fig.\ref{fig:Switch}(b). 
 

We can find the continuous-density solution for $P(\bold{m},t)$ in the large-ensemble limit by substituting the WKB ansatz into Eq.(\ref{eq:MasterEquation}), Taylor expanding Eq.(\ref{eq:MasterEquation}) in powers of the small parameter $1/C$, and neglecting terms of $\mathcal{O}(1/C)$ or smaller (see SI.\ref{sec:1} for expansion). As is customary, this approximation converts the master equation into a familiar Hamilton-Jacobi equation (HJE) from analytical mechanics \cite{Dykman1994,AssafRev}:
\begin{align}
\label{eq:HJE}
\frac{\partial S}{\partial t}+H(\bold{m},\partial S/\partial \bold{m})=0, 
\end{align} 
where $S$ and $H$ are called the Action and Hamiltonian, respectively \cite{AssafRev}. Once $S$ is found from Eq.(\ref{eq:HJE}), the distribution of large fluctuations is determined.
Following analytical mechanics, we define a momentum, $p_{i}\!=\!\partial S/\partial m_{i}$, and conveniently express the Hamiltonian,
\begin{align}
\!\!H(\bold{m},\bold{p})\!=\!\sum_{i}\!\!\Bigg[\frac{1}{2}(1\!-\!m_{i})(e^{2p_{i}}\!-\!1)\!\Bigg(\!\!\frac{1}{1\!+\!e^{-2\beta\!\sum_{j}\!A_{ij}m_{j}}}\!+f_{i}\!\!\Bigg)+\frac{1}{2}(1\!+\!m_{i})(e^{-2p_{i}}\!-\!1)\!\Bigg(\!\!\frac{1}{1\!+\!e^{2\beta\!\sum_{j}\!A_{ij}m_{j}}}\!+f_{i}\!\!\Bigg)\!\Bigg]\!.
\label{eq:Hamiltonian}
\end{align}   
A crucial result of the WKB approximation, which makes the approach worthwhile, is that solutions of the HJE extremize $S$, when expressed as the integral\cite{Dykman1994}
\begin{align}
S(\bold{m},t)=\int_{t_{0}}^{t}{\Big[\bold{p}\cdot\frac{d\bold{m}}{dt'}-H(\bold{m},\bold{p})\Big]dt'}=\int_{\bold{m}\small{(t=t_{0})}}^{\bold{m}}\!\!\bold{p}\cdot{d\bold{m}}-\int_{t_{0}}^{t}H(\bold{m},\bold{p})dt',
\label{eq:Action}
\end{align} 
where $\bold{m}(t)$ and $\bold{p}(t)$ are determined from Hamilton's equations of motion below, Eqs.(\ref{eq:EOM1}-\ref{eq:EOM2}). Since $S$ is minimized, the probability of a stochastic path associated with $S$ is maximized. In summary, when the transition between ordered states in a network is exponentially rare, there is a {\it least-action} path in $(\bold{m},\bold{p})$ phase-space that connects the two, which is a local maximum in probability, and therefore corresponds to the OP through a network. The Action along the OP gives $\rho(\bold{m},t)$ from Eq.(\ref{eq:Distribution}) and Eq.(\ref{eq:Action}).      

Finally, just as in analytical mechanics, a convenient approach for computing the OP is to solve Hamilton's equations of motion for the system, $\partial H/\partial p_{i}=\dot{m_{i}}$ and $\partial H/\partial m_{i}=-\dot{p_{i}}$:
\begin{align}
\label{eq:EOM1}
\dot{m}_{i}=&\frac{(1\!-m_{i})e^{2p_{i}}}{1\!+e^{-2\beta\!\sum_{j}A_{ij}m_{j}}}-\frac{(1\!+m_{i})e^{-2p_{i}}}{1\!+e^{2\beta\!\sum_{j}A_{ij}m_{j}}}+f_{i}\big[(1-m_{i})e^{2p_{i}}-(1+m_{i})e^{-2p_{i}}\big], \\
\label{eq:EOM2}
\dot{p}_{i}=&\frac{\frac{1}{2}\!(e^{2p_{i}}-\!1)}{1\!+e^{-2\beta\!\sum_{j}A_{ij}m_{j}}}-\frac{\frac{1}{2}\!(e^{-2p_{i}}-\!1)}{1\!+e^{2\beta\!\sum_{j}A_{ij}m_{j}}} - \beta\sum_{j}A_{ji}\!\Bigg[\!\frac{(1\!-m_{j})\!(e^{2p_{j}}-\!1)-(1\!+m_{j})\!(e^{-2p_{j}}-\!1)}{\big(\!e^{\beta\!\sum_{k}A_{jk}m_{k}}\!+e^{-\beta\!\sum_{k}A_{jk}m_{k}}\!\big)^{2}}\!\Bigg]+\frac{f_{i}}{2}\big[e^{2p_{i}}-e^{-2p_{i}}\big]. 
\end{align}
It is important to notice that if one takes $\bold{p}\!\equiv\!\bold{0}$ in Eqs.(\ref{eq:EOM1}-\ref{eq:EOM2}) the ``quenched mean field" equations are derived for a kinetic Ising model as a special case which ignores fluctuations \cite{Lynn2016,Goltsev2012}. Similar findings have been shown recently for epidemic dynamics \cite{Hindes2017PRE}. Therefore, the OP formalism naturally generalizes deterministic approaches for network dynamics to include large fluctuations. In practice, all that is needed to find the OP are appropriate boundary conditions for Eqs.(\ref{eq:EOM1}-\ref{eq:EOM2}). These are derivable directly from the distribution as explained in Sec.\ref{sec:BC}.

\subsection{\label{sec:BC} Optimal paths, distribution, and switching times}
By considering histograms of $\bold{m}(t)$ from time-series data, e.g., Fig.\ref{fig:Switch}(a), we can see that the expected exponential distribution of large fluctuations appears. Moreover, simple inspection gives us the boundary conditions for solving Eqs.(\ref{eq:EOM1}-\ref{eq:EOM2}). Since the distribution takes a maximum value at the equilibrium $\bold{m}^*$ (satisfying $\dot{\bold{m}}\!=\!\dot{\bold{p}}\!=\bold{0}$), the initial boundary condition for a large fluctuation is $\bold{m}(t\!=\!0)\!=\!\bold{m}^*$ and $\partial S\!/\!\partial\bold{m}\!=\bold{0}\!=\!\bold{p}(t\!=\!0)$. Because we are interested in large fluctuations that lead to a switch, the final boundary condition is similarly $\bold{m}(t\!\rightarrow\!\infty)\!=\!-\bold{m}^*$ and $\bold{p}(t\!\rightarrow\!\infty)\!=\!\bold{0}$ \cite{Dykman1994,AssafRev,Lindley2013}. Therefore, by solving Eqs.(\ref{eq:EOM1}-\ref{eq:EOM2}) subject to {\it zero-momentum boundary conditions}, we determine the OP: a single trajectory that gives the probability of a large fluctuation to $\bold{m}$ within logarithmic accuracy, $\ln{\rho(\bold{m})} \approx -S(\bold{m})$. We note that since the distribution is nearly constant in time, the network Action is time-independent $\partial S/\partial t\!=0\!=\!H\;\forall t$. Therefore, $S(\bold{m})$ is equal to the line integral of the momentum along the OP, from Eq.(\ref{eq:Action}).  

\begin{figure}[ht]
\centering
\includegraphics[scale=0.33]{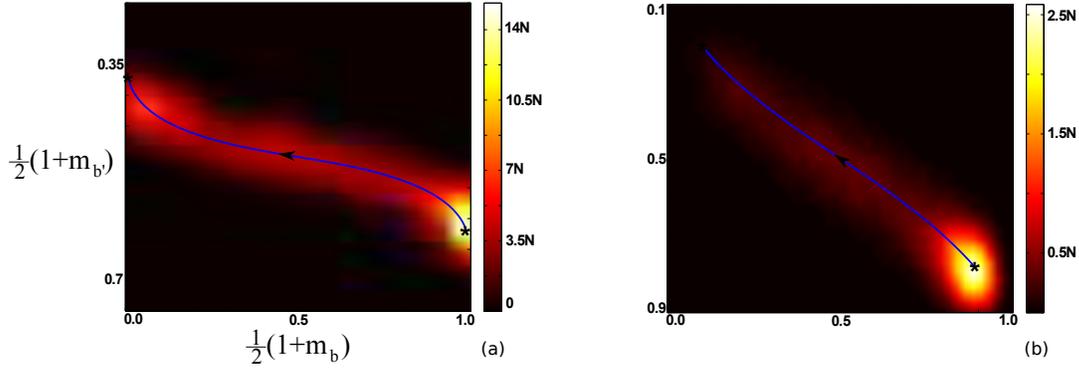}
\caption{Prehistory heat maps showing the density of the final $N$ stochastic events in 400 stochastic simulations that resulted in switching for two networks. Trajectories are projected into two network bins, denoted by subscripts $b$ and $b'$. (a) Network with a power-law degree distribution and parameters $\beta\lambda\!=\!1.625\!$ and $f_{i}\!=\!0\;\forall i$. The the x and y axes denote, respectively: the fraction of nodes in state $s_{i}\!=\!1$ with degrees $70\!\leq\!k_{i}\!\leq\!235$ and $10\!\leq\!k_{i}\!\leq\!12$. (b) Corresponding heat map for a Facebook network with parameters $\beta\lambda\!=\!1.37\!$ and $f_{i}\!=\!0.02\;\forall i$. The the x and y axes denote, respectively: the fraction of nodes in state $s_{i}\!=\!1$ with eigenvector centrality $\eta_{i}\!\geq\!0.0802$ and $0.0383\!\leq\!\eta_{i}\!\leq\!0.0580$. Network details can be found in SI.\ref{sec:3}. Arrows indicate the direction in time.}
\label{fig:HeatMaps}
\end{figure}

\begin{figure}[t]
\centering
\includegraphics[scale=0.32]{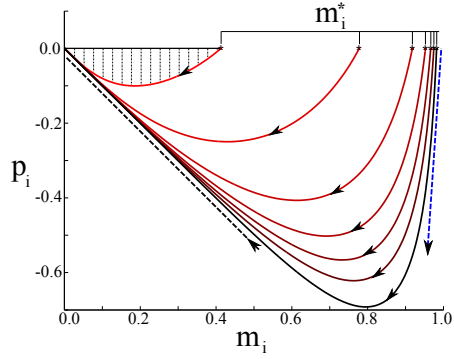}
\caption{Momentum versus opinion density for several Facebook network eigenvector centralities. The centralities are (from red to black) $\eta_{i}\!=\!0.0058, 0.0384, 0.0582, 0.0685, 0.0748, 0.0802,$ and $0.0864$. The shaded area illustrates the contribution to the Action for the lowest centrality. Analytic scalings for large fluctuations near $\bold{m}\!\approx\!\bold{0}$ and $\bold{m}\!\approx\!\bold{m}^{*}\!\lesssim\!\bold{1}$ are shown with black and blue dashed lines, respectively. Arrows indicate the direction in time. Parameters are $\beta\lambda\!=\!2.0\!$ and $f_{i}\!=\!0.02\; \forall i$.}
\label{fig:Momenta}
\end{figure}

Optimal paths can be computed numerically from Eqs.(\ref{eq:EOM1}-\ref{eq:EOM2}) with zero-momentum boundary conditions using quasi-newton methods\cite{Lindley2013} and dimension-reduction techniques based on a spectral decomposition of $A$ (see SI.\ref{sec:6}-\ref{sec:7} for details on numerical approaches). In general, we find two distinct segments of the OP: an activation segment with $\bold{p}\!\leq0$, requiring noise to carry the network from $\bold{m}\!=\!\bold{m}^{*}$ to $\bold{m}\!=\!\bold{0}$ (shown in Fig.\ref{fig:Momenta}), and a deterministic segment with $\bold{p}\!=\!\bold{0}$ that leads from $\bold{m}\!=\!0$ to $\bold{m}\!=\!-\bold{m}^{*}$. Since the deterministic segment has $\bold{p}\!=\!\bold{0}$, the probability distribution is predicted to be flat from $\bold{m}\!=\!0$ to $\bold{m}\!=\!-\bold{m}^{*}$, as demonstrated in Fig.\ref{fig:Switch}(b). In addition to the distribution, the shapes of large fluctuations are predicted and can be compared to simulations. For example, Fig.\ref{fig:HeatMaps} shows OP projections and prehistory trajectory-density plots from many stochastic realizations of Eq.(\ref{eq:Flip}) for two networks. We see that the OP for each network lies near the maximum of the corresponding heat maps. Importantly, since the OP predicts the most likely sequence of changes in opinion density, $\bold{m}$-projections at various points along a path (such as in Fig.\ref{fig:HeatMaps}) give us insight into how and where noise acts in a network during a large fluctuation.

Beyond finding a prescription for computing the OP numerically, we are interested in understanding how the path structure is related to topological properties of a network. The OP dependencies can be found analytically in certain limiting cases of interest, such as near threshold. A general method used throughout Sec.\ref{sec:BC} is to expand Eqs.(\ref{eq:EOM1}-\ref{eq:EOM2}) around the OP boundaries -- giving the local functional forms of $\bold{m}$, $\bold{p}$, and $S(\bold{m})$. We show in SI.\ref{sec:4} that when $\bold{f}\!=\!\bold{0}$ and the distance to threshold is small, $\delta\!\equiv\!\beta\lambda\!-\!1\!\gtrsim\!0$ (where $\lambda$ is the largest eigenvalue of $A$), the OP depends on the principle right and left eigenvectors of $A$, $\boldsymbol{\eta}$ and $\boldsymbol{\zeta}$, respectively:
\begin{align}
\label{eq:Path1}
\bold{m}(h)=&\boldsymbol{\eta}\delta^{1/2}h\sqrt{3/\sum_{j}\zeta_{j}\eta_{j}^{3}}, \\
\label{eq:Path2}
\bold{p}(h)=&\boldsymbol{\zeta}\delta^{3/2}h\big(h-1\big)\!\big(h+1\big)\!\sqrt{3/\sum_{j}\zeta_{j}\eta_{j}^{3}}.
\end{align}
$h$ is a unit-length parameter along the activation segment, $h\!\equiv\!m_{i}/m_{i}^{*}\;\forall i$. In general, the principle right eigenvector satisfies $\boldsymbol{\eta}\!=\!A\boldsymbol{\eta}\!/\!\lambda$, and similarly $\boldsymbol{\zeta}\!=\boldsymbol{\zeta}A\!/\!\lambda$ with $\boldsymbol{\zeta}\!\cdot\!\boldsymbol{\eta}\!=\!1$; $\eta_{i}$ is called the right eigenvector centrality of node $i$, and is an approximate measure of the importance of node $i$ in the network\cite{Newman:Book}. Near threshold $\bold{m}$ and $\bold{p}$ are parallel to the eigenvectors of $A$. Therefore, if $\boldsymbol{\eta}$ and $\boldsymbol{\zeta}$ contain relatively few nodes that have significantly large eigenvector centrality compared to most others, we expect the opinion density and fluctuations to be similarly large at such nodes. Further examining Eqs.(\ref{eq:Path1}-\ref{eq:Path2}), we see that the momentum is a cubic function of $h$ to lowest order in $\delta$, implying that $\ln{\rho(\bold{m})}$ has a curvature with respect to opinion density that is twice as large at $\bold{m}^{*}$ compared to $\bold{m}\!=\!\bold{0}$, and therefore the probability changes more quickly near $\bold{m}^{*}$ than near $\bold{m}\!=\!\bold{0}$. This can be contrasted with large fluctuations that cause extinction in epidemics, for which the momentum is linear\cite{HindesExtin2016}. 

In general, since each node's contribution to the Action is equal to the line integral of the momentum, we expect the contribution to increase with increasing eigenvector centrality. This pattern can be seen in Fig.\ref{fig:Momenta}, where the area under the darker curves (corresponding to higher centrality) contains the area under the lighter curves. Near threshold, we can calculate the line integral explicitly by substituting Eqs.(\ref{eq:Path1}-\ref{eq:Path2}) into Eq.(\ref{eq:Action}): $S(\bold{m})\!=\!\int_{\bold{m}*}^{\bold{m}}\bold{p}\cdot\bold{dm}'\!\approx\!\sum_{i=1}^{i=N}\!\int_{1}^{h}p_{i}(h')[dm_{i}/dh']dh'$ or    
\begin{align}
\label{eq:SmallAction}
S(\bold{m}(h))=&\frac{3\delta^{2}}{4\sum_{j}\zeta_{j}\eta_{j}^{3}}\big(1-h^{2}\big)^{2}. 
\end{align}
The expression is interesting, since for a symmetric network we find that $S$ is a function of the fourth moment of the eigenvector-centrality distribution, or inverse participation ratio \cite{Pastor2016}. To understand what this implies we first consider the case of a homogeneous complete graph, where $\eta_{i}\!=\!\zeta_{i}\!=\!1/\sqrt{N}.$ In this case the probability of a large fluctuation to zero consensus, and therefore the probability of switching, is to logarithmic accuracy $\ln{\rho(\bold{0})}\!=-3N\delta^2\!/4$ (for $N\delta^{2}\!\gg\!1$). On the other hand, for a random network without degree correlations the standard annealed-network approximation gives $\eta_{i}\!=\!\zeta_{i}\!=\!k_{i}/\!\sqrt{N\left<k^{2}\right>},$ given a degree $k_{i}$ for node $i$ and a network average of $k^{2}$, $\left<k^{2}\right>$. Hence, $\ln{\rho(\bold{0})}\!=-\!3N\delta^2\!\left<k^{2}\right>^{2}\!\!\!\big/4\left<k^{4}\right>$. If the the network is composed of nodes with only degree $k$, the switching probability is predicted to equal the complete graph's value. However for a heterogeneous network, e.g., when the distribution of nodes with degree $k$ scales like $k^{-\gamma}$, the additional topological factor, $\left<k^{2}\right>^{2}\!\!\!\big/\!\left<k^{4}\right>$, can be significantly smaller than $1$ depending on the cutoff of the distribution. Generally then, given the same distance to threshold and system size the switching probability is predicted to increase exponentially with increasing topological heterogeneity \cite{HindesExtin2016}. This will be particularly the case if $\gamma\!\leq\!5$, which can be contrasted to other kinetic systems such as epidemics \cite{Hindes2017PRE}.  

Near threshold a network is highly disordered, and we may wonder how large fluctuations behave as the order increases with $\beta$ far above threshold. In fact, the scaling patterns for the very largest fluctuations to small $\bold{m}\!\approx\!\bold{0}$ maintain a similar form. In SI.\ref{sec:5} we show that $m_{i}$ and $p_{i}$ remain proportional to centrality in the tail of the distribution by expanding Eqs.(\ref{eq:EOM1}-\ref{eq:EOM2}) around the origin (assuming $f_{i}\!=\!f$ and $A$ is symmetric). Hence, the relative probabilities for observing small ordering are simple functions of a network's average-squared opinion density, $\left<m^{2}\right>\!=\!(\bold{m}\cdot\bold{m})/\!N\gtrsim0:$
\begin{align}
\label{eq:SmallMajority}
\rho(\bold{m})/\rho(\bold{m}')=\exp{\!\!\bigg[\frac{N}{2}\!\bigg(\!\frac{\beta\lambda-1-2f}{1+2f}\!\bigg)\!\!\Big(\!\!\!\left<m^{2}\right>-\left<m'^{2}\right>\!\!\!\Big)\!\!\bigg]}. 
\end{align}
The generic scaling with $\left<m^{2}\right>$ occurs because $p_{i}$ tends to the same slope with respect to $m_{i}$ along the OP for all nodes when $\bold{m}\!\approx\!\bold{0}$; the scaling is demonstrated in Fig.\ref{fig:Momenta} with a black-dashed line. Notably Eq.(\ref{eq:SmallMajority}) implies that the tail of $\rho(\bold{m})$ is an exponential with a rate that increases {\it linearly} with the network size and distance to threshold, $\beta\lambda-1-2f$, but decreases with $f$. The latter effect entails an additional broadening of the distribution in the presence of spontaneous flipping that is independent of network topology. 

In contrast, the scaling at high order is significantly different, with fluctuations that are very sensitive to node centrality. In particular, when a network is near consensus, $\bold{m}\!\approx\!\bold{m}^{*}\!\lesssim\!\bold{1}$, lower-centrality nodes have fluctuations that are {\it exponentially} larger than higher-centrality nodes along the OP, $[m_{i}-m_{i}^{*}]\!/\![m_{j}-m_{j}^{*}]\sim[\eta_{i}/\eta_{j}]e^{2\beta\lambda\sum_{l}\!\eta_{l}m_{l}^{*}[\eta_{j}-\eta_{i}]},$ and are distributed according to a simple Gaussian:
\begin{align}
\label{eq:LargeOrder}
\rho(\bold{m})=b\exp{\!\!\bigg[\!\sum_{i}-\frac{1}{8}(m_{i}-m_{i}^{*})^{2}e^{2\beta\lambda\mathlarger{\eta_{i}\sum_{l}\!\eta_{l}m_{l}^{*}}}\!\bigg]}. 
\end{align}  
Intuitively, when a network is near consensus, it is very improbable for high centrality nodes to change state. This pattern can be seen in Fig.\ref{fig:Momenta} where the slope of $p_{i}$ with respect to $m_{i}$ along the OP is much steeper for higher centrality nodes when $\bold{m}\!\approx\!\bold{m}^{*}$-- corresponding to a more rapid decrease in probability from the maximum value. The quantitative scaling for high order is depicted in Fig.\ref{fig:Momenta} with a blue-dashed line, and reflects the pattern that the standard deviation of each node's distribution is exponentially decreasing with its centrality near consensus. Equation (\ref{eq:LargeOrder}) results from an expansion of Eqs.(\ref{eq:EOM1}-\ref{eq:EOM2}) near consensus, given in SI.\ref{sec:5}, and assumes $f\!\approx\!0$ and $A\!\approx\!\lambda\boldsymbol{\eta}\boldsymbol{\eta}^{T}$.

The final observable we consider in this report is the average switching time, $\left<T\right>$, since it quantifies the expected time scale over which the very largest fluctuations occur. Generally $\left<T\right>$ takes the form: 
\begin{align}
\label{eq:Time}
\left<T\right>\!=\!\tau(\beta,f,A)e^{S(\bold{0})}, 
\end{align}
from the assumption that switching between ordered states has a rate, or inverse time, proportional to the probability, $\rho(\bold{0})$ \cite{AssafRev,Assaf2010}. For sufficiently large $S$, the exponential contribution dominates, and therefore $\ln{\left<T\right>}\!\sim\!S(\bold{0})$, as demonstrated in Fig.\ref{fig:SwitchTimes}(a) for several networks. Given the dependence on $\rho(\bold{0})$, our analysis indicates that switching times have exponential sensitivity to the topological properties which determine the Action, such as network heterogeneity. 

\subsection{\label{sec:Control} Action-minimization and control}
Since we are able to predict large fluctuations with the OP theory, we may be interested in adding controls to the network dynamics in order to optimize an observable-- e.g., minimize $\left<T\right>$. In this report, our approach will involve minimizing $S(\bold{0})$ (a deterministic quantity derived from theory). However, it is important to note that the formalism ultimately entails using internal fluctuations in the network to do work, which would not be possible in the absence of noise \cite{HindesExtin2016}. In this section we study the example of targeting a subset, $F,$ of nodes with spontaneous flipping, for which $f_{i}\!=f\!\neq0$ given $i\in F$, while all other nodes have rate equal to zero. We consider several cases using the Facebook network\cite{Mcauley2012} as an illustration, and answer the following questions: what nodes should be targeted in order to minimize $\left<T\right>$ (given a fixed number for targeting, $|F|$, and a fixed flipping rate, $f$), and whether targeting a larger subset of nodes at a lower rate, or a smaller set with a higher rate tends to minimize $\left<T\right>$.       

\begin{figure}[t]
\centering
\includegraphics[scale=0.306]{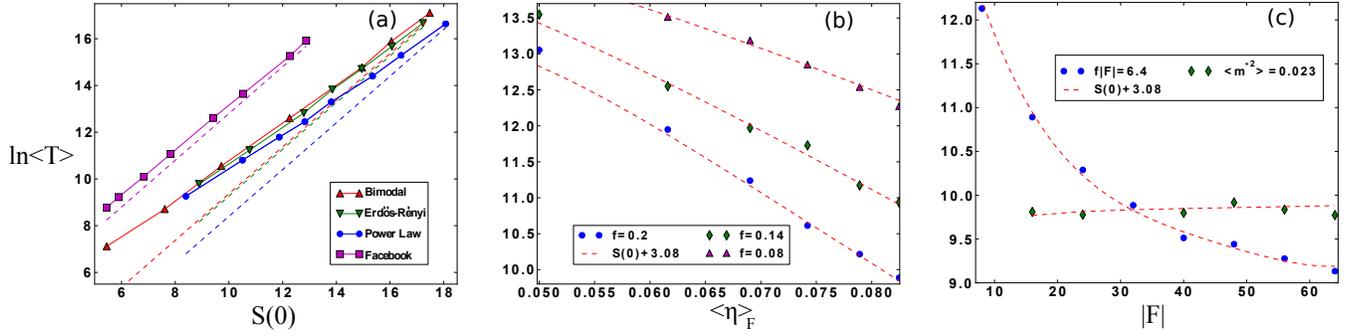}
\caption{(a) Log of the average switching times vs. Action, Eq.(\ref{eq:Action}), for several networks detailed in SI.\ref{sec:3}. $\left<T\right>$ is taken from at least $400$ stochastic realizations of Eq.(\ref{eq:Flip}) on a single fixed network, and shown with symbols. Dashed lines show the expected scaling $\ln{\!\left<T\right>}\!\approx\!S(\bold{0})\!+\!\text{constant}$. (b) Control of Facebook network for several flipping rates where the number of nodes targeted is constant, $|F|\!=\!32$, but the centrality of the targeted nodes is varied. $\left<\eta\right>_{\!F}$ is the average of $\eta_{i}$ in $F$. (c) Control of Facebook network where $|F|$ nodes with the highest $\eta_{i}$ are targeted, and the total control rate (blue circles) and amount of order (green diamonds) are held constant.}
\label{fig:SwitchTimes}
\end{figure}

Following the analysis above, we differentiate among nodes by eigenvector centrality, $\eta_{i}$ for node $i$. In order to first compare equal numbers of nodes for targeting, we have arranged nodes in a list according to increasing $\eta_{i}$, and binned the list into roughly equally sized bins. In Fig.\ref{fig:SwitchTimes}(b) $\ln{\!\left<T\right>}$ is shown as a function of the average $\eta_{i}$ in $F$, $\left<\eta\right>_{\!F}$, for several $f$. The size of $F$ is fixed at 32 nodes. The $F$ with highest $\left<\eta\right>_{\!F}$ in Fig.\ref{fig:SwitchTimes}(b) represents nodes near the maximum value of $\eta_{i}$ in the network: i.e., the $32$ nodes with highest centrality. The second highest $\left<\eta\right>_{\!F}$ represents nodes with the next highest centrality, and so on (see SI.\ref{sec:7}-\ref{sec:8} for more details). Since the times decrease with $\left<\eta\right>_{\!F}$, we can see that it is optimal to target nodes with higher centrality. The finding makes intuitive sense, since high $\eta_{i}$ implies high importance for a given node \cite{Newman:Book}, and therefore one might expect increased effect from flipping high-centrality nodes. Nevertheless, the scale of difference is important. Because of the exponential form of $\left<T\right>$, targeting nodes with the highest centrality can reduce $\ln{\!\left<T\right>}$ by $25\%$ for the parameters shown, even though $\!<\!1\%$ of the network is controlled. Predictions are in good agreement with simulations, Fig.\ref{fig:SwitchTimes}(b). 

On the other hand, with a different control scheme targeting higher $\eta_{i}$ alone may be sub-optimal. Another approach is to start with the 32 highest-centrality nodes (i.e., the control with the smallest $\ln{\!\left<T\right>}$ in Fig.\ref{fig:SwitchTimes}(b)), and increase/decrease the size of $F$ by adding/removing nodes with lower $\eta_{i}$. An example is shown in Fig.\ref{fig:SwitchTimes}(c) (blue circles) where $|F|$ nodes with the highest $\eta_{i}$ are targeted. In order to keep the total rate of flipping constant, $f|F|$ -- a proxy for the amount of work done by the controller, $f$ must vary accordingly. We can see that for the Facebook network it is more optimal to target a larger set of nodes at a slower rate, than fewer nodes at a higher rate, even though a larger set implies decreasing $\left<\eta\right>_{\!F}$. Similar to Fig.\ref{fig:SwitchTimes}(b), a reduction by a factor of $25\%$ in $\ln{\!\left<T\right>}$ is predicted and observed, despite targeting at most $\!<\!2\%$ of the network. Finally, it is interesting that minimizing $S(\bold{0})$, with the controls considered, is strongly correlated with reducing the amount of order in the network. To demonstrate, we redo the second control by picking an $f$ so that $\left<{m^*}^{2}\right>$ is constant as we vary $F$. The result is a slowly varying $\left<T\right>$, illustrated in Fig.\ref{fig:SwitchTimes}(c) (green diamonds).    


\section{Discussion}
There is much interest in understanding the relationship between dynamics in complex systems and their underlying topology. However, most theoretical results pertain to deterministic limits, where noise is ignored. Therefore, analytical and computational tools are needed to understand how noise and dynamics interact, especially when the interplay causes a large qualitative change in a network's collective dynamics. In this report, we have developed an approach based on WKB techniques applied to finite networks, which allowed us to analyze large fluctuations in order-disorder transitions driven by internal noise. The work went well beyond steady state and threshold analysis -- concerning itself with a {\it global} dynamical object, the optimal path of large fluctuations through a network. By computing the optimal path we were able to predict the probabilities and time scales associated with large fluctuations, including the largest fluctuations that entailed a noise-induced switch between deterministically stable states. Simulations in both real and synthetic networks showed good agreement with theory. 

The optimal path approach applied to networks has several advantages, which were detailed in this report. First, it provides a way to predict large fluctuations that naturally reproduces popular mean-field results as a special {\it zero-momentum limit}. Because of this structure, we expect more accurate network-approximation techniques, such as those that include dynamical correlations, to be generalized in a similar way -- and for many other dynamical processes in networks. For instance, WKB techniques can be used to analyze large fluctuations in nonlinear systems with time delays\cite{Schwartz2015PRE}, colored noise\cite{KamenevPRL2008}, and memory effects\cite{Bovenkamp2015}. Second, it allows one to analytically quantify the scaling patterns of large fluctuations on topology, including the exponential sensitivity of fluctuation-probabilities to topological heterogeneity and the multi-step structure of large fluctuations from highly ordered states through heterogeneous networks. Third, optimal control of noisy network dynamics is reduced to deterministic control of a mechanical analog. By minimizing the network {\it Action} derived from the optimal path theory, one can construct the optimal way to leverage internal noise so as to maximize the probability of network switching. This was demonstrated in a Facebook network where minimum-Action controls correctly predicted large exponential reductions in the average switching time, by targeting an optimal subset of the network that represented less than two percent of the total. Fourth, we expect that our approach will be useful for current avenues of research in network science and many new applications, such as network inference from data in the presence of large fluctuations.  

\bibliography{sample}

\section{Acknowledgements}
J. H. is a National Research Council postdoctoral fellows. I.B.S was supported by the U.S. Naval Research Laboratory funding (N0001414WX00023) and office of Naval Research (N0001416WX00657) and (N0001416WX01643). We are very grateful to L. B. Shaw and L. Mier-y-Teran-Romero for useful discussions. 

\section{Author contributions statement}
Both J.H. and I.B.S contributed to the methods and analysis, and wrote the manuscript.  

\section{Additional Information}
{\bf Supplementary information} accompanies this paper\\ 
\noindent {\bf Competing financial interests:} The authors declare no competing financial interests. 

\appendix

\section{Master equation expansion\label{sec:1}}
We want to find the leading contribution to Eq.(\ref{eq:MasterEquation}) when $C\!\gg\!1$, where $C$ is the number of stochastic realizations of the dynamical process defined by Eq.(\ref{eq:Flip}). Taylor expanding the probability and rates we find 
\begin{align}
\label{eq:1}
P(\bold{m}\!\pm\!\frac{2}{C}\bold{1}_i,t)=ae^{-CS(\bold{m}\pm\frac{2}{C}\bold{1}_{i},t)}\approx ae^{-C\big[S(\bold{m},t)\pm\frac{2}{C}\frac{\partial S}{\partial m_{i}}(\bold{m},t)+...\big]}=ae^{-CS(\bold{m},t)}e^{\mp2p_{i}},
\end{align}
\begin{align}
\label{eq:2}
R^{+}_{i}(\bold{m}\!-\!\frac{2}{C}\bold{1}_i)\approx R^{+}_{i}(\bold{m})+\Bigg[\frac{1}{1+e^{-2\beta\sum_{j}A_{ij}m_{j}}}\!+f_{i}\Bigg], 
\end{align}
and    
\begin{align}
\label{eq:3}
R^{-}_{i}(\bold{m}\!+\!\frac{2}{C}\bold{1}_i)\approx R^{-}_{i}(\bold{m})+\Bigg[\frac{1}{1+e^{2\beta\sum_{j}A_{ij}m_{j}}}\!+f_{i}\Bigg]. 
\end{align}
Substituting these approximations into Eq.(\ref{eq:MasterEquation}) gives: 
\begin{align}
\label{eq:4}
&-C\frac{\partial S}{\partial t}ae^{-CS(\bold{m},t)}=Cae^{-CS(\bold{m},t)}H(\bold{x},\bold{p})\nonumber \\ 
&\sum_{i}ae^{-CS(\bold{m},t)}e^{2p_{i}}\!\Bigg[\!\frac{1}{1+e^{-2\beta\sum_{j}A_{ij}m_{j}}}\!+f_{i}\!\Bigg]+ae^{-CS(\bold{m},t)}e^{-2p_{i}}\!\Bigg[\!\frac{1}{1+e^{2\beta\sum_{j}A_{ij}m_{j}}}\!+f_{i}\!\Bigg]
\end{align}
where $H(\bold{m},\bold{p})$ is given by Eq.(\ref{eq:Hamiltonian}). Dividing by $C$ and $ae^{-CS(\bold{m},t)}$, and neglecting the $\mathcal{O}(1/C)$ sum, we find Eq.(\ref{eq:HJE}). 

\section{Equilibria and linear spectra\label{sec:2}}
\label{sec:Spectra}
In general Eqs.(\ref{eq:EOM1}-\ref{eq:EOM2}) have three equilibria, $\bold{m}\!=\!\bold{0}\;\text{and}\;\pm\bold{m}^{*}$, that satisfy $\dot{\bold{m}}\!=\!\dot{\bold{p}}\!=\!\bold{0}$ with $\bold{p}\!=\!\bold{0}.$ In the special case where the spontaneous flipping rate is homogeneous, $f_{i}\!=\!f\; \forall i$, we find a set of fixed-point conditions:
\begin{align}
\label{eq:5}
m_{i}^{*}=\frac{\tanh\!\big(\beta\sum_{j}A_{ij}m_{j}^{*}\big)}{1+2f}. 
\end{align}
Often, it is convenient to approximate $A$ by the largest term in its eigenvalue decomposition, i.e., a large spectral-gap assumption. When $A$ is symmetric, $A\!=\!A^{T}\!\approx\!\lambda\boldsymbol{\eta\eta}^{T}$, where $\eta_{i}$ is the eigenvector centrality of node $i$ and $\lambda$ is the largest eigenvalue. Hence, a single equation determines the ordered equilibria (states) in terms of the order-parameter, $\overline{m}^{*}\!=\!\sum_{i}\!\eta_{i}m_{i}^{*}\!/\!\sum_{j}\!\eta_{j}$: 
\begin{align}
\label{eq:6}
\overline{m}^{*}=\frac{\sum_{i}\!\eta_{i}\tanh{\!\big(\beta\lambda\overline{m}^{*}\!\eta_{i}\!\sum_{j}\!\eta_{j}\big)}}{[1+2f]\!\sum_{l}\!\eta_{l}}, 
\end{align}
where $\sum_{i}\!\eta_{i}^{2}\!=\!1$. The linear stability spectra of the equilibria are found by substituting $\bold{m}\!=\!\bold{m}^{*}\!+\!\boldsymbol{\epsilon}$ and $\bold{p}\!=\!\boldsymbol{\mu}$ into Eqs.(\ref{eq:EOM1}-\ref{eq:EOM2}) and solving to $\mathcal{O}(\epsilon)$ and $\mathcal{O}(\mu)$. Because the resulting equations are linear, the dynamics are exponential: $\bold{\epsilon}(t)\!=\!\bold{\epsilon}e^{\sigma t}$ and $\bold{\mu}(t)\!=\!\bold{\mu}e^{\sigma t}$. The linearized dynamics gives three equations for the exponent $\sigma$ and the relative size (shape) of the modes $\epsilon_{i}$ and $\mu_{i}$:   
\begin{align}
\label{eq:7}
\sigma(\bold{m}^{*})=1+2f-\beta\lambda\!\sum_{j}\!\eta_{j}^{2}\big[1-(1+2f)^{2}{m_{j}^{*}}^{2}\big],
\end{align}
\begin{align}
\label{eq:8}
\mu_{i}=M\eta_{i},
\end{align}
and 
\begin{align}
\label{eq:9}
\epsilon_{i}=\frac{\mu_{i}(1+2f)}{(\sigma+1+2f)}\!\Big[2(1-{m_{i}^{*}}^{2})+\frac{\beta\lambda}{\sigma}\big[1-(1+2f)^{2}{m_{i}^{*}}^{2}\big]\!\sum_{j}\!\eta_{j}^{2}[1-{m_{j}^{*}}^{2}]\Big],  
\end{align}   
where $M$ is an arbitrary constant. 

Ordered states emerge at a threshold where the equilibrium ($\bold{m}\!=\!\bold{0}$, $\bold{p}\!=\!\bold{0}$) changes stability, $\sigma(\bold{0})\!=\!0$:
\begin{align}
\label{eq:10}
\beta\lambda=1+2f.
\end{align}
When $\beta\lambda\!>\!1+2f$, ordered sates have $\sigma(\bold{m}^{*})\!>\!0$ and $\sigma(\bold{0})\!<\!0$ for the modes Eqs.(\ref{eq:8}-\ref{eq:9}). However, the mean-field assumptions implicit in our approach can be quantitatively inaccurate for the threshold depending on the network. However, the WKB approach can be augmented to include pairwise correlations, for example, which generally improves accuracy. We mention that other solutions are possible with $\mu_{i}\!\equiv\!0$, which have oppositely signed spectra in terms of stability, i.e., where $\bold{m}\!=\!0$ is unstable and $\bold{m}\!=\!\bold{m}^{*}$ is stable. In general, taking $\bold{p}\!\equiv\!\bold{0} \; \forall t$ in Eqs.(\ref{eq:EOM1}-\ref{eq:EOM2}) gives the so called ``quenched mean field" equations corresponding to a dynamic Ising model with random flipping. In this way, the condition $\bold{p}\!\neq\!\bold{0}$ is what allows a trajectory (i.e, the OP) to exist from a ``stable" to ``unstable" state in the deterministic mean-field theory. 
\section{Network details\label{sec:3}}
\label{sec:Networks}
The Facebook network used throughout the paper was taken from \href{url}{http://snap.stanford.edu/data/egonets-Facebook.html}. It contains 4039 nodes and 88234 edges. The power-law network in Fig.\ref{fig:HeatMaps}(a) was generated from the configuration model with degree(k) distribution, $g_{k}\!=\!k^{-2.5}\!/\!\sum_{k'=10}^{300}k'^{-2.5}$, and 600 nodes. In Fig.\ref{fig:SwitchTimes}(a) the Erd\H{o}s-R\'{e}nyi network had 500 nodes and 15000 edges and the bimodal network was generated from the configuration model with 400 nodes and two degree classes: 40 nodes had degree 50 and 360 nodes had degree 5.

\section{Near threshold OP\label{sec:4}}
As in the main text, we consider the special case where $f\!=\!0$ near threshold, with $\delta\!=\!\beta\lambda-1\!\gtrsim\!0$. Our approach is to find $\bold{m}^{*}$ and the linear dynamics (Eqs.(\ref{eq:7}-\ref{eq:9})) near $\bold{m}\!\approx\!\bold{0}$ and $\bold{m}\!\approx\!\bold{m}^{*}$ to lowest order in $\delta$. This will give us boundary conditions which can be used to determine the OP to the same order in $\delta$. Once the OP is found, we can explicitly perform the line integral of momentum that gives the Action, Eq.(\ref{eq:Action}), and hence the probability exponent in the distribution of large fluctuations, Eq.(\ref{eq:Distribution}). For this section, it is {\it not assumed that A has a large spectral gap nor is symmetric}.
 
First, we expand Eq.(\ref{eq:5}) in powers of $\delta$, $m_{i}\!=\!\delta^{1/2}m_{i,1}+\delta^{3/2}m_{i,2}+...$, and collect terms of the same order in $\delta$: 
\begin{align}
\label{eq:11}
\mathcal{O}(\delta^{1/2})&: \;\;\;\;\;  m_{i,1}=\frac{1}{\lambda}\sum_{j}A_{ij}m_{j,1}, \\
\label{eq:12}
\mathcal{O}(\delta^{3/2})&: \;\;\;\;\;  m_{i,2}=m_{i,1}-\frac{1}{3}m_{i,1}^{3}+\frac{1}{\lambda}\sum_{j}A_{ij}m_{j,2}.  
\end{align}
Eq.(\ref{eq:11}) implies $m_{i,1}\!=\!E\eta_{i}$, and taking the inner product of Eq.(\ref{eq:12}) with the left eigenvector, $\zeta_{i}$, corresponding to $\eta_{i}$, gives $E\!=\!\sqrt{3/\sum_{j}\zeta_{j}\eta_{j}^{3}},$ or 
\begin{align}
\label{eq:13}
m_{i}^{*}=\eta_{i}\delta^{1/2}\sqrt{3/\sum_{j}\zeta_{j}\eta_{j}^{3}}+\mathcal{O}(\delta^{3/2}). 
\end{align}
Note: $\sum_{i}\zeta_{i}\eta_{i}\!=\!1$. 

Next, we find the dynamics near $\bold{m}\!\approx\!\bold{0}$. Analogous to Eqs.(\ref{eq:7}-\ref{eq:9}) (without the symmetric $A$ assumption), 
\begin{align}
\label{eq:14}
-\sigma(\bold{0})\mu_{i}(\bold{0})=-\mu_{i}(\bold{0})+\frac{1+\delta}{\lambda}\sum_{j}A_{ji}\mu_{j}(\bold{0}),\\
\label{eq:15}
\sigma(\bold{0})\epsilon_{i}(\bold{0})=-\epsilon_{i}(\bold{0})+2\mu_{i}(\bold{0})+\frac{1+\delta}{\lambda}\sum_{j}A_{ij}\epsilon_{j}(\bold{0}).
\end{align}
Solving Eqs.(\ref{eq:14}-\ref{eq:15}) we find that $\epsilon_{i}\!=\!\mathcal{E}\eta_{i}$ and $\mu_{i}\!=\!-\delta\mathcal{E}\zeta_{i}$, which means that
\begin{align}
\label{eq:16}
\frac{\mu_{i}(\bold{0})}{\epsilon_{i}(\bold{0})}=\frac{dp_{i}}{dm_{i}}(\bold{0})= -\frac{\delta\zeta_{i}}{\eta_{i}},
\end{align}
or the derivative at the boundary $\bold{m}\!=\!\bold{0}$ and $\bold{p}\!=\!\bold{0}$. Similarly, we seek the derivative at the boundary $\bold{m}\!=\!\bold{m}^{*}$ and $\bold{p}\!=\!\bold{0}$ to the same order, $\mathcal{O}(\delta)$. The linearized dynamics are 
 \begin{align}
\label{eq:17}
&-\sigma(\bold{m}^{*})\mu_{i}(\bold{m}^{*})=-\mu_{i}(\bold{m}^{*})+\frac{1+\delta}{\lambda}\!\sum_{j}\!A_{ji}\mu_{j}(\bold{m}^{*})-\frac{3\delta}{\lambda\sum_{l}\!\zeta_{l}\eta_{l}^{3}}\!\sum_{j}\!A_{ji}\mu_{j}(\bold{m}^{*})\eta_{j}^{2}+\mathcal{O}(\delta^{2}),\\
\label{eq:18}
&\sigma(\bold{m}^{*})\epsilon_{i}(\bold{m}^{*})=-\epsilon_{i}(\bold{m}^{*})+\Bigg[2-\frac{3\delta\eta_{i}^{2}}{\sum_{l}\!\zeta_{l}\eta_{l}^{3}}\Bigg]\mu_{i}(\bold{m}^{*})+
\frac{1+\delta}{\lambda}\!\sum_{j}\!A_{ji}\epsilon_{j}(\bold{m}^{*})-\frac{3\delta\eta_{i}^{2}}{\lambda\sum_{l}\!\zeta_{l}\eta_{l}^{3}}\!\sum_{j}\!A_{ji}\epsilon_{j}(\bold{m}^{*})\nonumber \\
&\;\;\;\;\;\;\;\;\;\;\;\;\;\;\;\;\;\;\;\;\;\;\;\;\;+\mathcal{O}(\delta^{2}),
\end{align}
Solving Eqs.(\ref{eq:17}-\ref{eq:18}) gives $\epsilon_{i}\!=\!\mathcal{A}\eta_{i}$ and $\mu_{i}\!=\!2\delta\mathcal{A}\zeta_{i}$, implying a derivative boundary condition
\begin{align}
\label{eq:19}
\frac{\mu_{i}(\bold{m}^{*})}{\epsilon_{i}(\bold{m}^{*})}=\frac{dp_{i}}{dm_{i}}(\bold{m}^{*})=\frac{2\delta\zeta_{i}}{\eta_{i}}. 
\end{align}

Finally, it is convenient to parameterize $m_{i}$ and $p_{i}$ in terms of a unit-length parameter $h$, such that $h\!\equiv\!m_{i}/m_{i}^{*}\;\forall i$. Note: when $h\!=\!1$ the network is ordered at $\bold{m}^{*}$, and when $h\!=\!0$ the network has no order $\bold{m}\!=\bold{0}$. Therefore we can write, $m_{i}(h)\!=\!m_{i}^{*}h$ and $p_{i}(h)\!=\!\delta m_{i}^{*}\zeta_{i}f(h)/\eta_{i}$, where $f(h)$ is an unknown function that we must determine. The boundary conditions above imply: $f(h\!=\!1)\!=\!0$, $f(h\!=\!0)\!=\!0$,  $\frac{df}{dh}(h\!=\!1)\!=\!2$, and $\frac{df}{dh}(h\!=\!0)\!=\!-1$. If we assume that $f(h)$ is a polynomial, the simplest polynomial that satisfies the four boundary conditions is a cubic function, $f(h)\!=\!h\big(h-1\big)\!\big(h+1\big)$. Hence we arrive at Eqs.(\ref{eq:Path1}-\ref{eq:SmallAction}). We mention that the near threshold OP is a convenient initial guess for the Iterative-Action-Minimization-Method described in Sec.\ref{sec:IAMM}.       


\section{Scaling away from threshold\label{sec:5}}
We would like to use Eqs.(\ref{eq:7}-\ref{eq:9}) to find basic scalings of the OP away from threshold, where the network's metastable order is high, which will help us understand fluctuations near $\bold{m}\!\approx\!\bold{0}$ and $\bold{m}\!\approx\!\bold{m}^{*}$ -- i.e., the largest and smallest fluctuations.  We first study the former for which the solution of Eqs.(\ref{eq:7}-\ref{eq:9}) is useful: $\mu_{i}/\epsilon_{i}\!=\![1+2f-\beta\lambda]/[1+2f]\!=\!\frac{dp_{i}}{dm_{i}}(\bold{0}).$ Therefore, the momentum is linear in $m_{i}$ with constant slope across the network, $p_{i}\!\approx\!m_{i}[1+2f-\beta\lambda]/[1+2f]$. By considering the action at $\bold{m}\!=\!\bold{0}$, $S(\bold{0})\!=\!\sum_{i}\!\int_{m_{i}}^{0}p_{i}(m_{i}^{'})dm_{i}^{'}+\sum_{i}\!\int_{m_{i}^{*}}^{m_{i}}p_{i}(m_{i}^{'})dm_{i}^{'}\!=\!\sum_{i}\!\int_{m_{i}}^{0}p_{i}(m_{i}^{'})dm_{i}^{'}+S(\bold{m}),$ or $-S(\bold{m})\!=\!-S(\bold{0})-\!\sum_{i}\!\int_{0}^{m_{i}}p_{i}(m_{i}^{'})dm_{i}^{'},$ we find
\begin{align}
\label{eq:20}
-S(\bold{m})\approx-S(\bold{0})+\Bigg[\frac{\beta\lambda-1-2f}{1+2f}\Bigg]\!\sum_{i}\!\int_{0}^{m_{i}}\!\!m_{i}^{'}dm_{i}^{'}=-S(\bold{0})+\Bigg[\frac{\beta\lambda-1-2f}{1+2f}\Bigg]\!\sum_{i}\!\frac{m_{i}^{2}}{2}. 
\end{align} 
Using Eq.(\ref{eq:19}) we derive the relative probabilities for the very largest fluctuations, i.e., the tail of the large-fluctuation distribution to small $\bold{m}$, or Eq.(\ref{eq:SmallMajority}). 

A similar technique gives the small fluctuations near global consensus, $\bold{m}\!\approx\!\bold{m}^{*}\!\lesssim\!\bold{1}$. Assuming $f\!\approx\!0$ and $A\!\approx\!\lambda\boldsymbol{\eta\eta}^{T}$, Eq.(\ref{eq:9}) gives $\epsilon_{i}/\epsilon_{j}\!=\![1-{m_{i}^{*}}^{2}]/[1-{m_{j}^{*}}^{2}]$. In this region, ${m_{i}^{*}}^{2}\!\approx\!1-4e^{-2\beta\lambda\eta_{i}\!\sum_{j}\!\eta_{j}m_{j}^{*}}$, and therefore $[m_{i}-m_{i}^{*}]/[m_{j}-m_{j}^{*}]\sim[\eta_{i}/\eta_{j}]e^{2\beta\lambda\sum_{l}\!\eta_{l}m_{l}^{*}[\eta_{j}-\eta_{i}]}$ -- showing that fluctuations for low-centrality nodes are exponentially larger than for high-centrality nodes. Moreover, combining with Eq.(\ref{eq:7}-\ref{eq:8}) we find $\sigma(\bold{m}^{*})\!\approx\!1$ and 
\begin{align}
\label{eq:21}
\frac{\mu_{i}(\bold{m}^{*})}{\epsilon_{i}(\bold{m}^{*})}=\frac{dp_{i}}{dm_{i}}(\bold{m}^{*})\approx\frac{1}{1-{m_{i}^{*}}^{2}}=\frac{1}{4}e^{2\beta\lambda\eta_{i}\sum_{l}\!\eta_{l}m_{l}^{*}}. 
\end{align} 
For $\bold{m}\!\approx\!\bold{m}^{*}$, $S(\bold{m})\!\approx\!\sum_{i}\int_{m_{i}^{*}}^{m_{i}}\frac{dp_{i}}{dm_{i}^{'}}(\bold{m}^{*})(m_{i}^{'}-m_{i}^{*})dm_{i}^{'}$, and thus 
\begin{align}
\label{eq:22}
S(\bold{m})\!\approx\!\sum_{i}-\frac{1}{8}(m_{i}-m_{i}^{*})^{2}e^{2\beta\lambda\eta_{i}\!\sum_{l}\!\eta_{l}m_{l}^{*}},
\end{align} 
which is equivalent to Eq.(\ref{eq:LargeOrder}). 

\section{\label{sec:IAMM}Finding the OP numerically\label{sec:6}}
In general, one would like to find the OP beyond the limiting cases. Of course, no analytic solution is possible except in networks that are effectively low-dimensional. 
Since the path connects two equilibria via a heteroclinic orbit, in practice it must be constructed numerically, e.g., through shooting, or quasi-newton methods, etc. The method used in this report is of the latter form, namely the Iterative-Action-Minimizing-Method (IAMM) (\href{url}{doi:10.1016/j.phys d.2013.04.001}). In this method, OPs are generated from a least-squares algorithm that minimizes the residuals between Eqs.(\ref{eq:EOM1}-\ref{eq:EOM2}) and finite-difference approximations, with fixed-point boundary conditions from Sec.\ref{sec:BC} (used to close the differencing). However the dimension for the minimization is $2Nd$ where $d$ is the number of discrete points in the differencing and $N$ is the size of the network, which is prohibitively large for large $N$ (typically we choose $200\!\leq\!d\!\leq\!500$). Therefore, in practice it is necessary to coarse-grain the network in some way. We describe our approach for this report in Sec.\ref{sec:bin}. We mention that for the special case of $\bold{f}\!=\!\bold{0}$, the OP is {\it reversible}, and therefore $d\bold{m}/dt$ along the first segment is time reversed along the second.

\section{Binning the network\label{sec:7}}
\label{sec:bin}
We are interested in reducing the dimension of network defined by the adjacency,
matrix, $A\in{\cal {L}}({\cal R}^{N},{\cal R}^{N})$. All of the networks considered in this report are symmetric,
though the formalism does not require this assumption. Nevertheless, in this section we assume $A\!=\!A^{T}$.  
Given a sequence of of eigenvalues, $\{\lambda_{j}\}$, and eigenvectors, $\{\bm{\eta}_{j}\},$
for $A,$ we assume the largest eigenvalue is much greater
than all of the others. This is a good approximation for many networks, including those discussed in Sec.\ref{sec:Networks}. 
From the spectral decomposition theorem, we
can approximate the adjacency matrix as $A\approx\lambda\bm{\eta\eta}^{T}$,
where $\lambda=max\{\lambda_{i}\}$, and $\bm{\eta}$ the corresponding
eigenvector.

In order to create a mapping from $N$ dimensions to one that is significantly
lower, we first notice that the entries of the eigenvector (which we assume
is normalized) roughly relate a measure of vertex importance in the
graph. Therefore, we reorder the entries of $\bm{\eta}$ in increasing
order such that $\bm{v\!=\!P\eta}$, where $\bm{P}\in{\cal {L}}({\cal R}^{N},{\cal R}^{N})$
is a permutation matrix, and $v_{1}\le v_{2}\le\dots\le v_{N}$. Notice
we have not changed the norm of $\bm{v}$, nor have we made any dimension
reduction.

Next, we arbitrarily pick a binning of $\bm{v}$ such that there exists
$|B|\!\ll\!N$ bins, and associated with each bin $b\in B$ we have
a distribution, $g_{b}$, as well as an index set, ${\cal I}_{b}$.
We define an indicator function on an index such $\chi_{b}(i)=1$
if $i\in{\cal I}_{b}$, 0 otherwise. We now define a vector that
averages the nodes within a bin $b$ as the following:

\begin{align}
r_{b} & =(1/Ng_{b})[\chi_{b}(1),\chi_{b}(2),\dots\chi_{b}(N)]\cdot\bm{v}\label{eq:Projection1}\\
\equiv & \bm{\alpha_{b}^{T}v}.
\end{align}

That is, $Ng_{b}$ is the total number of nodes in bin $b,$
and $r_{b}=(1/Ng_{b})\sum_{i\in{\cal I}_{b}}v_{i}$.

The map in Eq.(\ref{eq:Projection1}) computes the average over all
of those nodes in bin $b.$ To compute the entire transformation
from ${\cal R}^{N}$into ${\cal R}^{B},$we have 

\[
\bm{r}=\left[\begin{array}{cccc}
\bm{\alpha}_{1}^{T} & \cdot & \cdot & \cdot\\
\bm{\alpha}_{2}^{T} & \cdot & \cdot & \cdot\\
\vdots & \cdot & \cdot & \cdot\\
\bm{\alpha}_{\beta}^{T} & \cdot & \cdot & \cdot
\end{array}\right]\bm{P\bm{\eta}\equiv\bm{{\cal A}P\eta},}
\]
where $\bm{{\cal A}\in}{\cal L}({\cal R}^{N},{\cal R}^{B})$. Using the same transformation map for $\bold{m}$ and $\bold{p}$, we find the corresponding $|B|$ dimensional vectors, $\boldsymbol{\mathcal M}$ and $\boldsymbol{\mathcal P}$, respectively, for the average opinion density and momentum in bins. By replacing $v_{i}$ with $r_{b}$, $m_{i}$ with $\mathcal{M}_{b}$, 
and $p_{i}$ with $\mathcal{P}_{b}$ for $i\in{\cal I}_{b}$ in Eqs.(\ref{eq:EOM1}-\ref{eq:EOM2}), we get the (approximate) equations of motion for bin $b$:  
\begin{align}
\label{eq:EOM3}
\dot{{\mathcal M}}_{b}=&\frac{(1\!-{\mathcal M}_{b})e^{2{\mathcal P}_{b}}}{1\!+e^{-2\beta\lambda r_{b}\!\sum_{b'}\!Ng_{b'}r_{b'}{\mathcal M}_{b'}}}-\frac{(1\!+{\mathcal M}_{b})e^{-2{\mathcal P}_{b}}}{1\!+e^{2\beta\lambda r_{b}\!\sum_{b'}\!Ng_{b'}r_{b'}{\mathcal M}_{b'}}}+f_{b}\big[(1-{\mathcal M}_{b})e^{2{\mathcal P}_{b}}-(1+{\mathcal M}_{b})e^{-2{\mathcal P}_{b}}\big], \\
\label{eq:EOM4}
\dot{{\mathcal P}}_{b}=&\frac{\frac{1}{2}\!(e^{2{\mathcal P}_{b}}-\!1)}{1\!+e^{-2\beta\lambda r_{b}\!\sum_{b'}\!Ng_{b'}r_{b'}{\mathcal M}_{b'}}}-\frac{\frac{1}{2}\!(e^{-2{\mathcal P}_{b}}-\!1)}{1\!+e^{2\beta\lambda r_{b}\!\sum_{b'}\!Ng_{b'}r_{b'}{\mathcal M}_{b'}}}+\frac{f_{b}}{2}\big[e^{2{\mathcal P}_{b}}-e^{-2{\mathcal P}_{b}}\big] \nonumber \\
&- \beta\lambda r_{b}\!\sum_{b'}\!Ng_{b'}r_{b'}\!\Bigg[\!\frac{(1\!-{\mathcal M}_{b'})\!(e^{2{\mathcal P}_{b'}}-\!1)-(1\!+{\mathcal M}_{b'})\!(e^{-2{\mathcal P}_{b'}}-\!1)}{\big(\!e^{\beta\lambda r_{b'}\!\sum_{b''}\!Ng_{b''}r_{b''}{\mathcal M}_{b''}}\!+e^{-\beta\lambda r_{b'}\!\sum_{b''}\!Ng_{b''}r_{b''}{\mathcal M}_{b''}}\!\big)^{2}}\!\Bigg],
\end{align} 
assuming $f_{i}\!=\!f_{b}$ $\forall$ $i\in{\cal I}_{b}$. A final requirement is needed to ensure that the binned and original system have the same bifurcation point and are similarly normalized: after binning we {\it renormalize} $r_{b}$ so that $\sum_{j}\eta_{j}^{2}\!=\sum_{b}r_{b}^{2}g_{b}N\!=\!1$.

\begin{figure}[h]
\centerline{\includegraphics[scale=0.37]{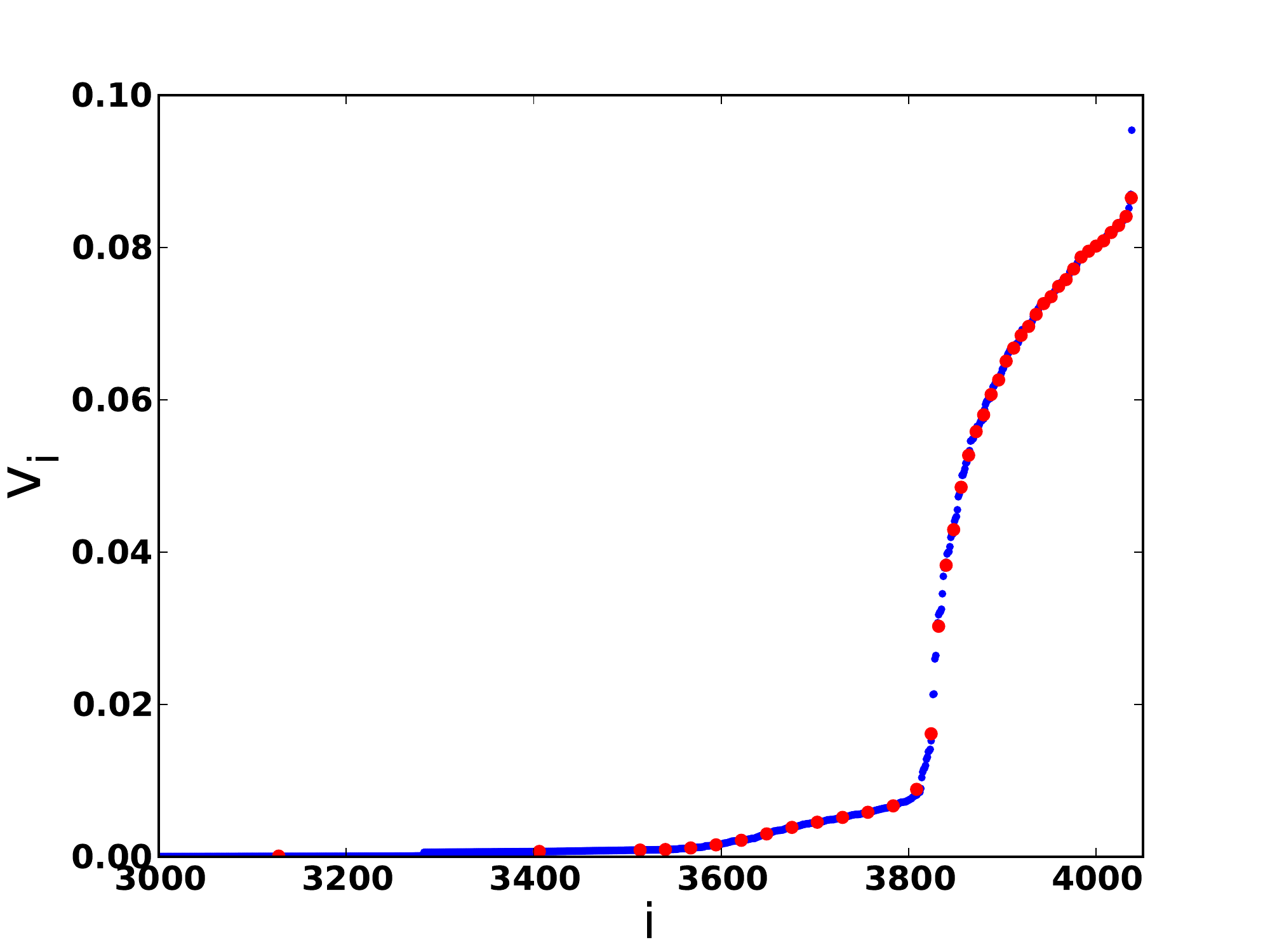}}
\caption{Example binning procedure for Facebook network. Eigenvector centralities (blue) are shown for all network positions and compared with the average centrality in each bin (red)}
\label{fig:Binning}
\end{figure}

In practice, to use the binning procedure one must specify $\chi_{b}(i)$. We illustrate with the Facebook network, where we chose $|B|\!=\!50$. The $v_{i}$ distribution is shown in Fig.\ref{fig:Binning} in blue. Note, the first $3000$ nodes have small $v_{i}$, and therefore we truncate the x-axis for easier viewing. Visually the $v_{i}$ has roughly three relevant parts: $v_{i}\sim$ $\mathcal{O}(0.1)$, $\mathcal{O}(0.01)$, and $\mathcal{O}(0.001)$ or smaller. The binned distribution is shown in red.
We chose to bin each of the three parts with roughly equal numbers of nodes in each bin -- with a total of $28$, $12$, and $10$ bins for the three parts, respectively. Such a choice gave the following indicator functions, $\chi_{b}(i)$, which we list in their entirety for completeness: 
\begin{equation*}
\begin{aligned}[c]
\chi_{1}(i)&=1 \;\text{if}\;  0.09541<\!v_{i}, \nonumber \;\;\;\;\;\; \\
\chi_{2}(i)&=1 \;\text{if}\;  0.08605<\!v_{i}\!\leq0.09541, \nonumber \;\;\;\;\;\;  \\
\chi_{3}(i)&=1 \;\text{if}\;  0.08352<\!v_{i}\!\leq0.08605, \nonumber \;\;\;\;\;\;  \\
\chi_{4}(i)&=1 \;\text{if}\;  0.08226<\!v_{i}\!\leq0.08352, \nonumber  \;\;\;\;\;\; \\
\chi_{5}(i)&=1 \;\text{if}\;  0.08155<\!v_{i}\!\leq0.08226, \nonumber  \;\;\;\;\;\; \\
\chi_{6}(i)&=1 \;\text{if}\;  0.08040<\!v_{i}\!\leq0.08155, \nonumber  \;\;\;\;\;\; \\
\chi_{7}(i)&=1 \;\text{if}\;  0.08001<\!v_{i}\!\leq0.08040, \nonumber  \;\;\;\;\;\; \\
\chi_{8}(i)&=1 \;\text{if}\;  0.07913<\!v_{i}\!\leq0.08001, \nonumber  \;\;\;\;\;\; \\
\chi_{9}(i)&=1 \;\text{if}\;  0.07800<\!v_{i}\!\leq0.07913, \nonumber  \;\;\;\;\;\; \\
\chi_{10}(i)&=1 \;\text{if}\; 0.07682<\!v_{i}\!\leq0.07800, \nonumber  \;\;\;\;\;\; \\
\chi_{11}(i)&=1 \;\text{if}\; 0.07555<\!v_{i}\!\leq0.07682, \nonumber  \;\;\;\;\;\; \\
\chi_{12}(i)&=1 \;\text{if}\; 0.07433<\!v_{i}\!\leq0.07555, \nonumber  \;\;\;\;\;\; \\
\chi_{13}(i)&=1 \;\text{if}\; 0.07287<\!v_{i}\!\leq0.07433, \nonumber  \;\;\;\;\;\; \\
\chi_{14}(i)&=1 \;\text{if}\; 0.07234<\!v_{i}\!\leq0.07287, \nonumber  \;\;\;\;\;\; \\
\chi_{15}(i)&=1 \;\text{if}\; 0.07038<\!v_{i}\!\leq0.07234, \nonumber  \;\;\;\;\;\; \\
\chi_{16}(i)&=1 \;\text{if}\; 0.06939<\!v_{i}\!\leq0.07038, \nonumber  \;\;\;\;\;\; \\
\chi_{17}(i)&=1 \;\text{if}\; 0.06743<\!v_{i}\!\leq0.06939, \nonumber  \;\;\;\;\;\; \\
\chi_{18}(i)&=1 \;\text{if}\; 0.06652<\!v_{i}\!\leq0.06743, \nonumber  \;\;\;\;\;\; \\
\chi_{19}(i)&=1 \;\text{if}\; 0.06404<\!v_{i}\!\leq0.06652, \nonumber  \;\;\;\;\;\; \\
\chi_{20}(i)&=1 \;\text{if}\; 0.06209<\!v_{i}\!\leq0.06404, \nonumber  \;\;\;\;\;\; \\
\chi_{21}(i)&=1 \;\text{if}\; 0.05991<\!v_{i}\!\leq0.06209, \nonumber  \;\;\;\;\;\; \\
\chi_{21}(i)&=1 \;\text{if}\; 0.05695<\!v_{i}\!\leq0.05991, \nonumber  \;\;\;\;\;\; \\
\chi_{22}(i)&=1 \;\text{if}\; 0.05484<\!v_{i}\!\leq0.05695, \nonumber  \;\;\;\;\;\; \\
\chi_{23}(i)&=1 \;\text{if}\; 0.05095<\!v_{i}\!\leq0.05484, \nonumber  \;\;\;\;\;\; \\
\chi_{24}(i)&=1 \;\text{if}\; 0.04556<\!v_{i}\!\leq0.05095, \nonumber  \;\;\;\;\;\; \\
\chi_{25}(i)&=1 \;\text{if}\; 0.04072<\!v_{i}\!\leq0.04556, \nonumber \;\;\;\;\;\; \\
\end{aligned}
\begin{aligned}[c]
\chi_{26}(i)&=1 \;\text{if}\; 0.03454<\!v_{i}\!\leq0.04072,\nonumber  \;\;\;\;\;\; \\
\chi_{27}(i)&=1 \;\text{if}\; 0.02598<\!v_{i}\!\leq0.03454,\nonumber  \;\;\;\;\;\; \\
\chi_{28}(i)&=1 \;\text{if}\; 0.01309<\!v_{i}\!\leq0.02598,\nonumber  \;\;\;\;\;\; \\
\chi_{29}(i)&=1 \;\text{if}\; 0.00722<\!v_{i}\!\leq0.01309,\nonumber  \;\;\;\;\;\; \\
\chi_{30}(i)&=1 \;\text{if}\; 0.00624<\!v_{i}\!\leq0.00722,\nonumber  \;\;\;\;\;\; \\
\chi_{31}(i)&=1 \;\text{if}\; 0.00556<\!v_{i}\!\leq0.00624,\nonumber  \;\;\;\;\;\; \\
\chi_{32}(i)&=1 \;\text{if}\; 0.00488<\!v_{i}\!\leq0.00556,\nonumber  \;\;\;\;\;\; \\
\chi_{33}(i)&=1 \;\text{if}\; 0.00430<\!v_{i}\!\leq0.00488,\nonumber  \;\;\;\;\;\; \\
\chi_{34}(i)&=1 \;\text{if}\; 0.00347<\!v_{i}\!\leq0.00430,\nonumber  \;\;\;\;\;\; \\
\chi_{35}(i)&=1 \;\text{if}\; 0.00242<\!v_{i}\!\leq0.00347,\nonumber  \;\;\;\;\;\; \\
\chi_{36}(i)&=1 \;\text{if}\; 0.00191<\!v_{i}\!\leq0.00242,\nonumber  \;\;\;\;\;\; \\
\chi_{37}(i)&=1 \;\text{if}\; 0.00130<\!v_{i}\!\leq0.00191,\nonumber  \;\;\;\;\;\; \\
\chi_{38}(i)&=1 \;\text{if}\; 0.00103<\!v_{i}\!\leq0.00130,\nonumber  \;\;\;\;\;\; \\
\chi_{39}(i)&=1 \;\text{if}\; 0.00091<\!v_{i}\!\leq0.00103,\nonumber  \;\;\;\;\;\; \\
\chi_{40}(i)&=1 \;\text{if}\; 0.00086<\!v_{i}\!\leq0.00091,\nonumber  \;\;\;\;\;\; \\\;
\chi_{41}(i)&=1 \;\text{if}\; 0.000610<\!v_{i}\!\leq0.00086,\nonumber  \;\;\;\;\;\; \\
\chi_{42}(i)&=1 \;\text{if}\; 1.452329E\!-\!5<\!v_{i}\!\leq0.000610,\nonumber  \;\;\;\;\;\; \\
\chi_{43}(i)&=1 \;\text{if}\; 3.145425E\!-\!6<\!v_{i}\!\leq1.452329E\!-\!5,\nonumber  \;\;\;\;\;\; \\
\chi_{44}(i)&=1 \;\text{if}\;1.9930885E\!-\!6<\!v_{i}\!\leq3.145425E\!-\!6,\nonumber  \;\;\;\;\;\; \\
\chi_{45}(i)&=1 \;\text{if}\;2.4888568E\!-\!7<\!v_{i}\!\leq1.9930885E\!-\!6,\nonumber  \;\;\;\;\;\; \\
\chi_{46}(i)&=1 \;\text{if}\;9.2757062E\!-\!08<\!v_{i}\!\leq2.4888568E\!-\!7,\nonumber  \;\;\;\;\;\; \\
\chi_{47}(i)&=1 \;\text{if}\;5.9389469E\!-\!08<\!v_{i}\!\leq9.2757062E\!-\!08,\nonumber  \;\;\;\;\;\; \\
\chi_{48}(i)&=1 \;\text{if}\;2.2296092E\!-\!08<\!v_{i}\!\leq5.9389469E\!-\!08,\nonumber  \;\;\;\;\;\; \\
\chi_{49}(i)&=1 \;\text{if}\;6.5813248E\!-\!10<\!v_{i}\!\leq2.2296092E\!-\!08,\nonumber  \;\;\;\;\;\; \\
\chi_{50}(i)&=1 \;\text{if}\;6.2455171E\!-\!14<\!v_{i}\!\leq6.5813248E\!-\!08,\nonumber  \;\;\;\;\;\; \\
&\!\!\!\!\!\!\!\!\text{and}\; \chi_{b}(i)=0 \;\text{otherwise}\;\forall b. \nonumber
\end{aligned}
\end{equation*}
Note, $g_{b}\!=\!\sum_{i}\chi_{b}(i)/N$. Similar binning procedures were used for all other networks, Sec.\ref{sec:Networks}. 

\section{Control\label{sec:8}}
For the first control (Sec.\ref{sec:Control}, 2nd paragraph, main text), the control set $F$ with the largest $\left<\eta\right>_{\!F}$ corresponds to the first $4$ bins (centered on red points) starting from the right in Fig.\ref{fig:Binning}. In particular, the control set contains $32$ nodes with the highest $v_{i}$, or the first 32 blue points starting from the right in Fig.\ref{fig:Binning}. The next control set with the second largest $\left<\eta\right>_{\!F}$ corresponds to bins $5\!-\!8$ starting from the right in Fig.\ref{fig:Binning} -- namely, the next $32$ nodes with highest centrality but less than the lowest centrality in the first set of 32 nodes. This pattern is continued for six different control sets and three flipping rates, $f$, Fig.\ref{fig:SwitchTimes}(b).

For the second control (Sec.\ref{sec:Control}, 3rd paragraph, main text), again we start with the $32$ nodes with highest $v_{i}$ as our control set, and then add/subtract nodes with lower/higher $v_{i}$. For example, moving one point to the left along the $|F|$ axis in Fig.\ref{fig:SwitchTimes}(c)(main text) to $|F|\!=\!24$, implies controlling $24$ nodes with the highest $v_{i}$. Moving, one point to the right implies controlling $40$ nodes with the highest $v_{i}$, and so forth. Since the size of the control set is changed, we change $f$ in order to keep a quantity constant. The two constants chosen for Fig.\ref{fig:SwitchTimes}(c) were $f|F|$ (blue points) and $\sum_{i}{m_{i}^{*}}^{2}$ (green diamonds). We found little change in the Action when the order was held constant. We mention that this is not always the case: in epidemics, minimizing the epidemic size does not imply minimizing the Action in general (see Ref.[15], main text).

\label{sec:control}

\end{document}